# "Pharos Night: Crown Pursuit": An AI-Native Deck-Building and Tactical Arena Game Design Based on Multi-Agent Systems

Ting-Chen Hsu*
School of Animation and Digital Arts, Communication University of China, tingchenhsu.ac@gmail.com

Jueyao Liu*
School of Animation and Digital Arts, Communication University of China, s191937762@163.com

Yanzi Zhou*
School of Animation and Digital Arts, Communication University of China, 2080565170@qq.com

Jiangxu Lin*
School of Animation and Digital Arts, Communication University of China, chinalinjiangxu@gmail.com

Haoyu Xu*
School of Animation and Digital Arts, Communication University of China, 503083340@qq.com

Yuwen Liu*
School of Animation and Digital Arts, Communication University of China, 2577584858@qq.com

Yanjia Liu*
School of Animation and Digital Arts, Communication University of China, mrsaltdove@qq.com

Bangjing Xu*
School of Animation and Digital Arts, Communication University of China, 625537668@qq.com

---

* Equal contribution. Ting-Chen Hsu is responsible for the project's overall design and programming; Jueyao Liu handles the design of core mechanics and some programming; Yanzi Zhou is in charge of character and key visual design; Jiangxu Lin handles illustration and UI design; Haoyu Xu is responsible for environment design; Yuwen Liu and Yanjia Liu handle animation production; and Bangjing Xu is responsible for music production.

With advancements in generative AI technology, an increasing number of researchers have begun exploring AI-native games in which gameplay rules are directly driven by generative AI. This paper presents "Pharos Night: Crown Pursuit," an AI-native deck-building and tactical arena game based on a multi-agent system. The game uses large language models to generate materials and cards, support NPC decision-making, and mediate natural-language interactions. During play, players collect materials, describe desired card effects in natural language, and choose whether to negotiate or fight with NPCs in the arena. To constrain model-generated outcomes, the system parses responses as structured JSON, constructs card effects from predefined mechanics, and maps qualitative effect levels to designer-specified numerical values. A small-scale playtest with 13 participants suggests that the system can provide strategically meaningful and engaging AI-driven gameplay, while also revealing challenges related to predictability, transparency, and player control. This work demonstrates the potential of multi-agent generative AI systems for creating more emergent digital game experiences.

# 1 Introduction

In recent years, driven by the rapid advancements of generative AI (Gen-AI) technologies in narrative generation [1], NPC control [2], and development assistance [3], AI-native games are emerging as an entirely new gaming paradigm. By driving core gameplay and mechanics through Gen-AI, AI-native games can create unique and unprecedented interactive experiences for players [4]. Currently, some researchers have explored the design of the games driven by Gen-AI—for instance, creating gaming experiences with dynamic narratives [5], or developing serious games for purposes such as psychological healing [5], mitigating dialect bias [6], or engineering education [7]. However, introducing generative AI into digital games—such as card games and competitive games—still faces numerous challenges. How to simultaneously ensure the balance of AI-generated elements, the intelligence of NPCs, and the enjoyability of the game is a critical issue in current research.

Therefore, we have designed "Pharos Night: Crown Pursuit"—an AI-native deck-building and tactical arena game driven by multi-agent. Within the game, generative AI serves to dynamically generate in-game elements, enable free-form card synthesis, and drive the behaviors and natural language interactions of intelligent NPCs. Our work provides a new perspective for AI driven gameplay design, demonstrating how the dynamic generation and intelligent reasoning capabilities supported by generative artificial intelligence can be integrated into the core loop of games. In addition, we have introduced a multi-agent architecture that combines project generation, card synthesis, balance control, and intelligent NPCs to form a game system that supports dynamic emergence. This work establishes a new paradigm for AI native games and explores the novel design possibilities of using generative AI as a gameplay driven foundation.

# 2 Methods and Game Concept

## 2.1 Motivation and High-Level Goals

Deck-building and tactical arena games traditionally rely on carefully authored cards, predefined character behaviors, and fixed rule spaces. While such structures support balance and strategic clarity, they also limit the degree to which players can improvise with the game system itself. Our design began from a central question: how can generative AI become part of the playable rule space, rather than merely a tool for producing narrative or visual content?

The overall goals of this work can be summarized at three levels. The first goal is to transform deck-building from the selection of predefined cards into an open-ended synthesis process. Players collect AI-generated materials and combine them with natural language "spells" to create new magic cards, which allows player intention to become part of the construction process. The second goal is to make NPCs function as social and strategic agents, enabling them to explore the world, negotiate, trade, fight, and respond to players through natural language interaction. As a result, player strategy extends beyond capability enhancement to include persuasion, deception, alliance-building, and information control. The third goal is to explore new emerging gameplay in AI-native game design. While AI enables flexible content generation, it also carries the potential to disrupt game systems. Therefore, we have designed a number of specialized architectures to ensure a coherent gameplay experience while preserving the unique sense of surprise and expressiveness inherent in AI-driven gameplay.

## 2.2 Conceptual Gameplay and Core Loop

In "Pharos Night: Crown Pursuit," players take on the role of a student at Pharos—a renowned magic academy in the region of Corinia—and participate in its annual championship game. All students enter a mystical realm to compete, and the sole winner shall be crowned with "KETER." Figure 1 illustrates the core loop of gameplay. After entering the scene, players need to constantly explore to collect materials by searching for treasure chests or defeating monsters. After collecting some materials, players can choose to freely create magic cards through the synthesis system to enhance their combat power. When players face NPCs, they can choose to initiate battles, negotiate, or engage in other forms of communication. If two characters enter a battle, they assemble their own magic cards to engage in combat. Losers will be eliminated, while those who survive to the end will win.

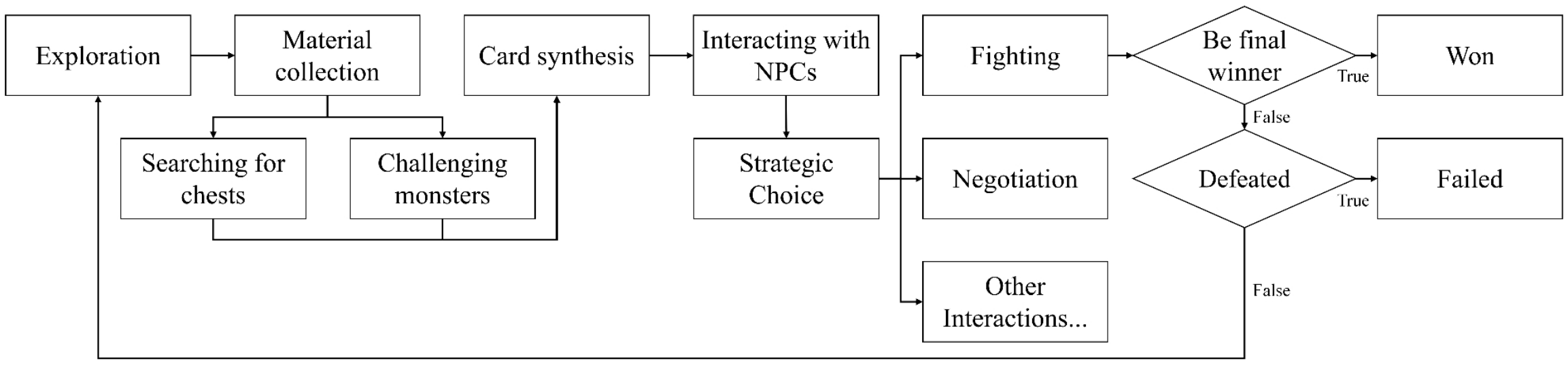

**Figure 1: The core loop of "Pharos Night: Crown Pursuit."**

## 2.3 Overview of Development Methods

Our project is developed using Unity. For mechanisms in the game that rely on LLMs, we use the ChatGLM series models from Zhipu for driving: predefined structured prompts are constructed in Unity, sent to the model service, and parsed back in Unity after receiving JSON responses. To manage model limitations and prevent excessive concurrent calls, we implement a request queue that allows at most three simultaneous large model invocations, with additional requests queued until slots free up.

For player testing, we use a combination of questionnaires and open-ended questions. The questionnaire employed a custom scale to gauge participants' understanding of the AI mechanisms, their perceived strategic meaningfulness and sense of control over them, as well as cognitive load and overall experience. Open-ended questions asked players to reflect on (1) moments when the AI- driven card synthesis felt strategically meaningful, (2) moments when the AI-driven NPCs bring a unique experience, and (3) suggestions for improving the AI mechanisms in the game.

# 3 GAME Design Results

## 3.1 The Module of Main Scene

The main scene allows players to explore freely (Figure 2. (a)), moving by right-clicking the ground. When the player moves the mouse over a specific interactive object, an interaction button will pop up. Treasure chests are one of the main ways for players to obtain materials (Figure 2. (b)). Once a treasure chest is opened, it will respawn at another random location on the map after 30 seconds. Each chest contains two materials, and some chests may also include an additional material with higher magic value. All materials are drawn from the material library created by the LLM at the start of the game. The LLM-driven material generation mechanism allows players to encounter new patterns each time they play the game. LLM-driven NPCs will roam the scene and may interact with players (Figure 2. (c)). Players can discuss tactics, form temporary cooperation, exchange information, and so on with them. Since NPCs' behavior is completely driven by LLMs, their interactions with players will be stored in their memories and further influence their behavior. NPCs can also interact or initiate combat with other NPCs. Free dialogue, such as with merchants (Figure 2. (d)) or special facilities (Figure 2. (e)), is also permitted. Unlike traditional games, players are free to express their needs and desires, and even bargain.

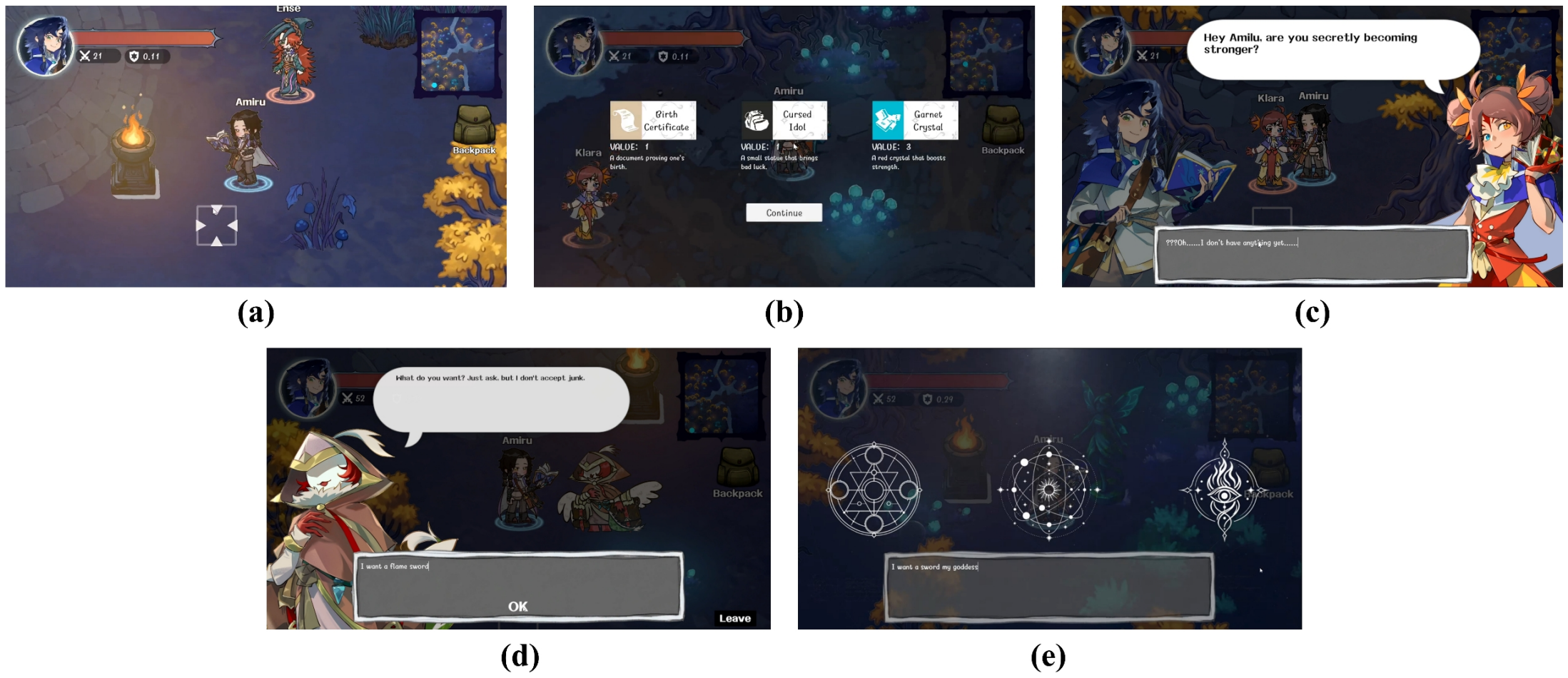

**Figure 2: The main scene module screens of the game. (a) Player freely explores the scene. (b) Player opens chests to obtain materials. (c) Player freely interacts with a NPC. (d) Player engages in a transaction. (e) Player interacts with the "Hermes Statue."**

## 3.2 Synthesis Mechanism

In the game, the various materials players acquire are stored in their inventory so they can use them in the synthesis system (Figure 3. (a)). Player can drags and drops materials to the synthesis area for synthesis (Figure 3. (b)). The upper right square of the synthesis area determines whether the magic card is a functional card or a combat card. The three middle squares allow players to put in any material. The input bar below allows players to enter spells to define the card effects they want. After the player places all the materials and clicks the synthesis button, the LLM will start calling the synthesis function to build cards. The properties of the materials in the middle three squares and the player's spell will determine the effect of the card, and the magic value of the three materials will determine the level of the card (each level represents a different strength). The degree to which LLM follows player spells is random, but if players make demands that are too out of range, it means that cards will have strong effects accompanied by severe negative effects. After the card is synthesized, it will be added to the player's backpack (Figure 3. (c)).

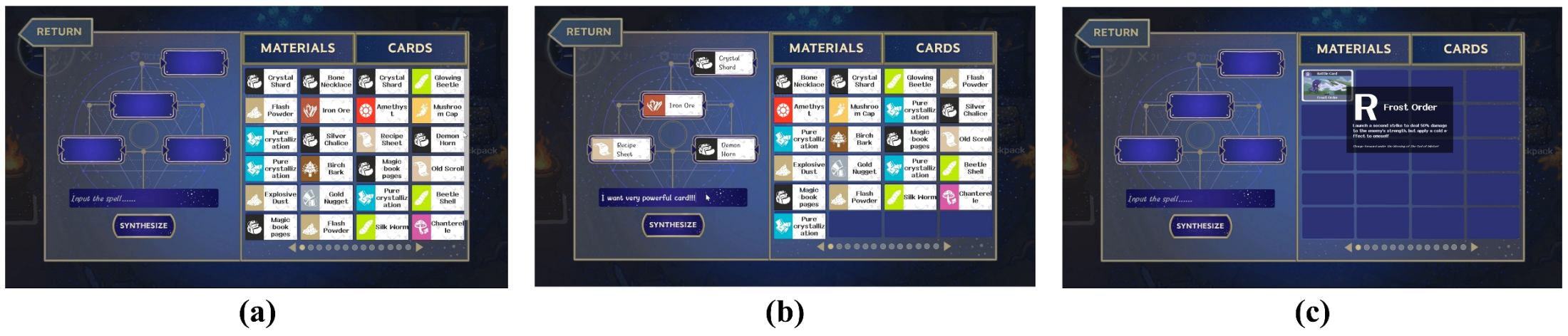

**Figure 3: The backpack and synthesis system screens of the game. (a) Player's backpack interface. (b) Player places materials in the synthesis area. (c) Players obtain magic cards through the synthesis system.**

## 3.3 The Module of Combat

During the interaction between the player and NPCs, combat can be initiated and transferred to the combat module at any time. The initiation of a combat can be unilateral, but initiating a combat without knowing the opponent's strength implies risks, as a ceasefire can only be reached with the agreement of both sides. A complete combat is divided into multiple rounds, and before each round begins, players can assemble their own magic cards or initiate a chat with their opponents (Figure 4. (a)). All cards assembled by both sides will be automatically executed in order during the combat (Figure 4. (b)). Automatic sequential execution means that players need to consider the special effects that opponents may apply to themselves (such as cold states), and pre-set countermeasure cards in the deck in advance (such as clearing negative states). After all the cards from both sides have been executed, the next round of preparation begins, during which free communication between the two characters is allowed (Figure 4. (c)). These communications allow for intimidation, surrender, or negotiation between characters, which may lead to a ceasefire or the surrender of opponents

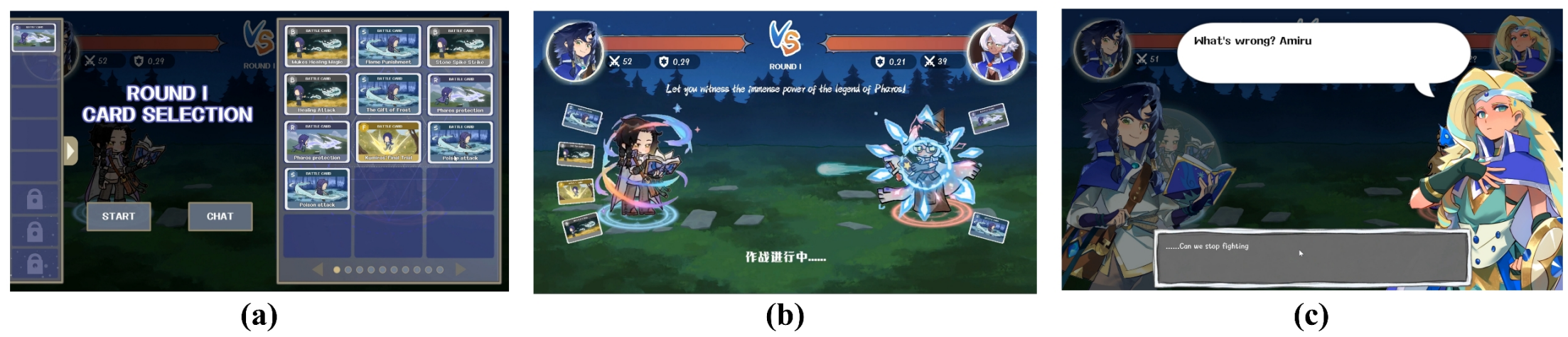

**Figure 4: The combat module screens of the game. (a) Combat preparation stage. (b) Combat in progress. (c) Inter-battle communication.**

# 4 System Implementation

## 4.1 Intelligent NPC system

Figure 5 shows the structure of NPCs in this game, which generally follows the process of "perception-thinking-decision-action". The default state of NPC is weighted roaming (prioritize going to unexplored places), and in this state, it calls the LLM for thinking every 5 seconds. Interruptions from external objects (such as interactions initiated by players or other NPCs) will terminate this state and shift towards waiting for external object actions and making decisions and actions. The special detection in the perceptron will also interrupt roaming state (such as detecting chests, characters, monsters). NPCs will turn to reading the world information from the perceptron and make further decisions or actions based on it. After any action is executed, it will default back to roaming mode. This cycle ensures that LLM-driven NPCs have independent personal memories and characteristic behaviors, while also ensuring that NPCs can make dynamic and natural responses based on time and environmental stimuli.

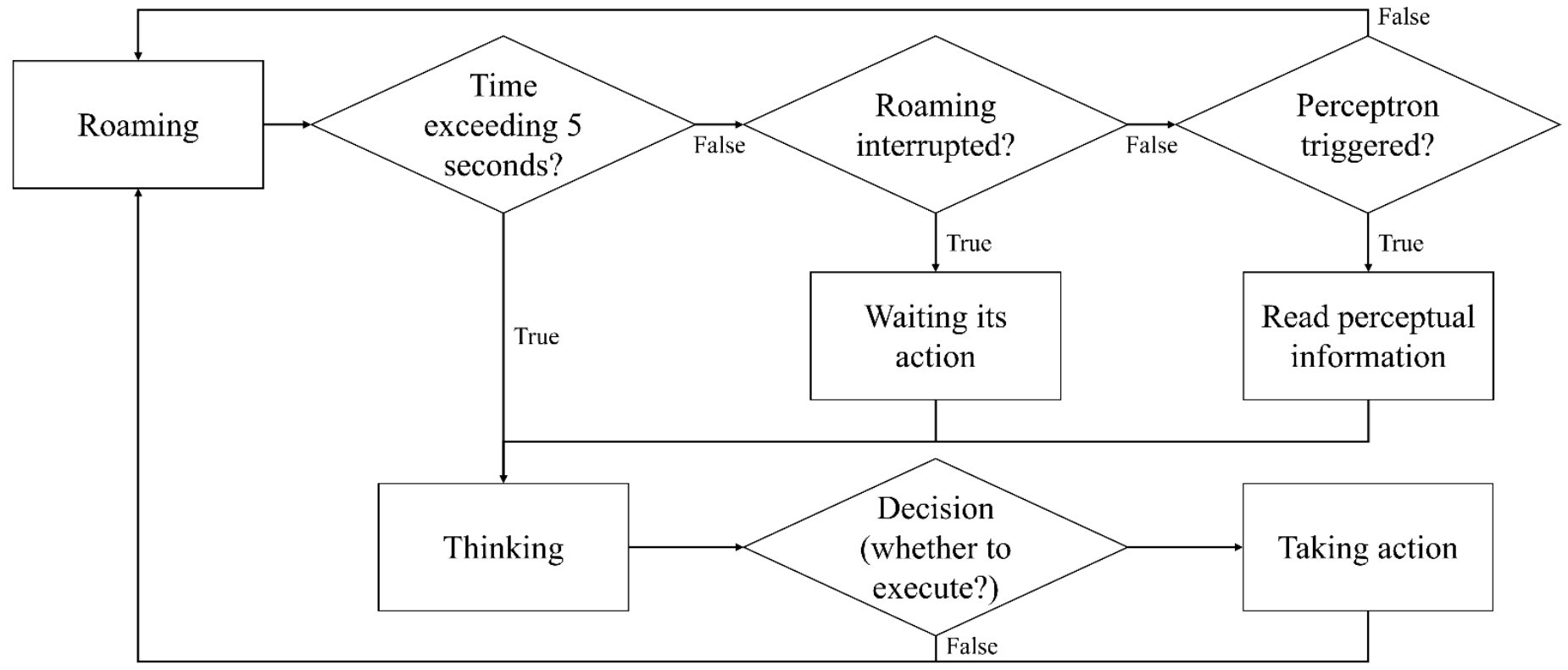

**Figure 5: Architecture design of intelligent NPCs in "Pharos Night: Crown Pursuit."**

## 4.2 Cards and Free Synthesis System

Figure 6 illustrates the process by which the LLM generates cards. The overall workflow encompasses information integration, JSON generation and parsing, numerical translation, and final card execution during combat.

The process begins with information integration. The Unity client aggregates four types of information into a prompt—output specifications, a list of meta-mechanics, placed prop assets, and player-input spells—and sends this to the LLM to generate the JSON.

Next, the LLM designs the card. Each card requires the definition of "pre-execution conditions," "pre-execution actions," "primary actions," "trigger conditions," and "trigger actions," all of which are designed by selecting and combining elements from the meta-mechanics list. The generated JSON is then returned to Unity for parsing, where Unity extracts each value and stores it in the corresponding variables of the new card object.

Regarding the specific numerical values for each card mechanism, the LLM generates qualitative descriptions across five levels (e.g., "low," "medium") rather than specific numbers. To ensure game balance, designers pre-define the specific numerical values corresponding to the five levels of each meta-mechanic; a numerical translation function then maps these qualitative levels to specific values stored in the card object's variables.

Finally, during the combat phase, each card executes in three stages, with each stage accessing the relevant variables of the card object. The process proceeds by evaluating "pre-execution conditions" and performing the associated actions, followed by the execution of the primary action mechanism, and concluding with the evaluation of "trigger conditions" and the execution of the corresponding actions.

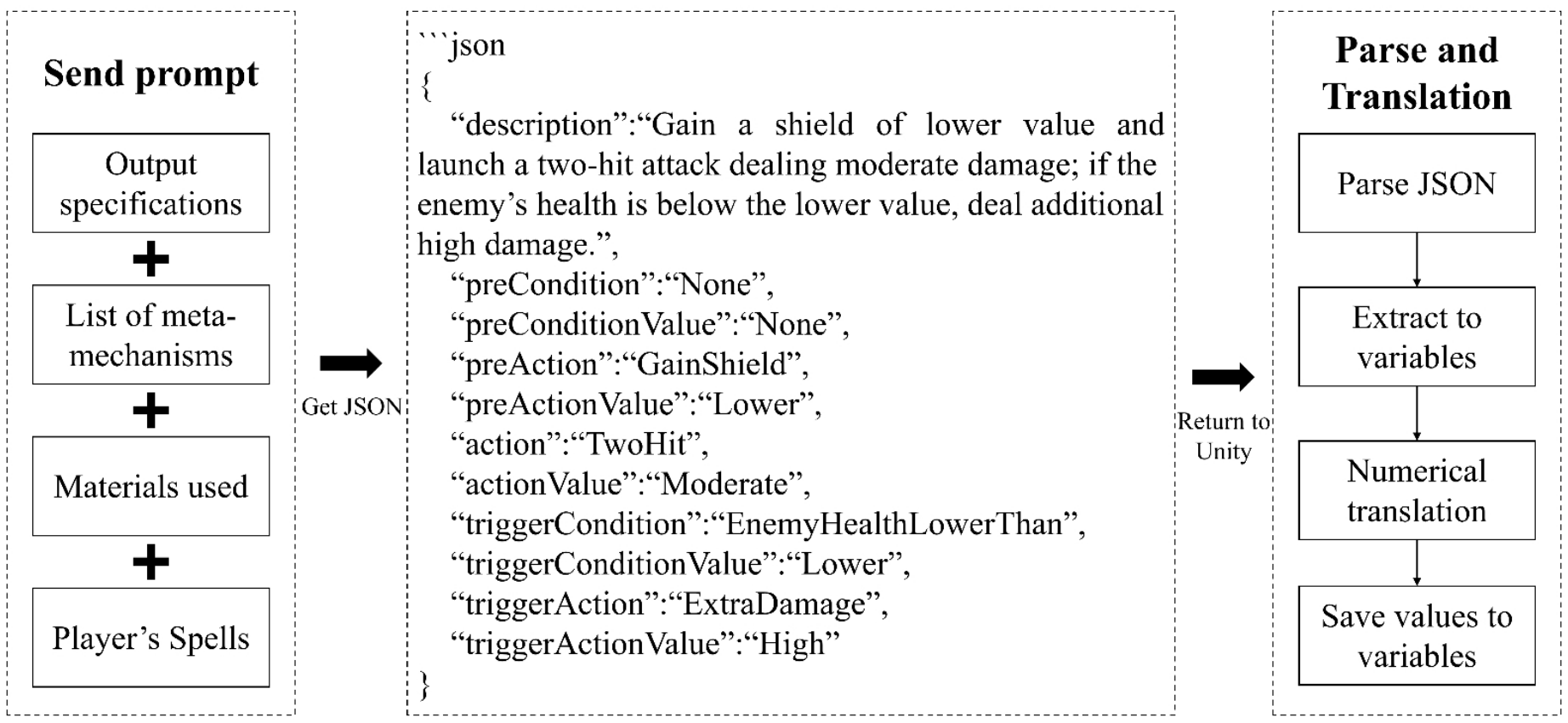

**Figure 6: Principles of LLM-based card generation.**

# 5 Playtesting Insights

To evaluate the player experience of this game, we recruited 13 players for a small-sample trial and validated the results using a questionnaire and open-ended questions.

## 5.1 Results of the Questionnaire Measurement

Table 1 shows the results of player experience based on questionnaire measurement (N=13). The measurement dimensions include understanding of the AI mechanisms, perceived strategic meaningfulness, sense of control, cognitive load, and overall experience (immersion, engagement and sense of ability). The questionnaire adopts a Likert five point scoring system.

**Table 1: Descriptive statistics of questionnaire results in various dimensions**

| | Understanding of the AI mechanisms | Perceived strategic meaningfulness | Sense of control | Cognitive load | Overall experience |
|---|---|---|---|---|---|
| Mean | 3.67 | 4.15 | 3.41 | 3.23 | 4.13 |
| S.D. | 0.62 | 0.48 | 0.72 | 0.75 | 0.62 |

Overall, players are positive about the AI-native gaming experience. The perceived strategic significance score is the highest (M=4.15, SD=0.48), indicating that players typically consider AI-driven card synthesis and NPC interaction as meaningful components of their strategic decisions. The overall experience also received positive feedback (M=4.13, SD=0.62), indicating that the game provides a relatively engaging and immersive experience.

Compared to these dimensions, players have a moderate understanding of AI mechanisms (M=3.67, SD=0.62), while their sense of control scores are slightly lower (M=3.41, SD=0.72). Therefore, although players appreciate the flexibility and novelty brought by artificial intelligence, some still find it difficult to fully predict or control the results generated by AI. The cognitive load remained at a moderate level (M=3.23, SD=0.75), indicating that the AI-driven mechanism did not impose too much burden, but there is still room for optimization. These preliminary results indicate that the system has successfully supported meaningful-AI driven games, while also revealing the need for clearer feedback and guidance mechanisms.

## 5.2 Results of the Open-ended Questions

The open-ended responses further reveals the attractiveness and limitations of AI-driven mechanisms. Many players believe that card synthesis is novel and significant (84.61%). Some people even reported that unexpected card mechanisms can create novel tactics (38.46%). Similarly, most players believe that AI-driven NPCs are more energetic and dynamic than traditional NPCs (92.30%). However, some players also pointed out that the high unpredictability of AI sometimes reduces their sense of control (46.15%). Therefore, some suggestions focus on improving transparency and guidance, such as providing more synthesis prompts, highlighting material features inherited by cards, and adding clearer guidance explanations. Overall, qualitative feedback indicates that AI mechanisms have successfully generated novelty and emergence, but improvements are still needed in terms of stability, predictability, and other aspects.

# 6 DISCUSSION AND CONCLUSION

This work presents AI-native game “Pharos Night: Crown Pursuit,” which integrates LLM-driven card synthesis, intelligent NPC behavior, and free-form player interaction into a gameplay loop. The playtesting results suggest that generative AI can meaningfully expand the interest and novelty of digital games. Players were able to influence card creation through materials and natural language prompts, while NPCs introduced negotiation, uncertainty, and emergent social dynamics beyond conventional combat-oriented play.

At the same time, our findings also reveal a central design issue in AI-native games. The same openness that creates novelty and emergence may also reduce predictability and sense of control. Future work should therefore focus on improving transparency and feedback, such as clearer synthesis hints, explanations of NPC decisions, and more stable balance constraints. Although the current study is limited by its small sample size, it demonstrates the potential of multi-agent systems as a foundation for more interesting and novel AI-driven gaming experiences.

## ACKNOWLEDGMENTS

This work did not receive funding. The author thanks four tutors, Shi Huang, Zhaogong Zhang, Jingwei Chen, and Ning Ding, for their guidance on the work. We thank all playtest participants for their time, effort and feedback. We are also grateful to the School of Animation and Digital Arts at Communication University of China for supporting the development and testing of this project.